\documentclass[journal]{IEEEtran}
\usepackage{graphicx}
\usepackage{caption2}
\usepackage{subfigure} 
\usepackage{float}
\usepackage{color}
\usepackage{amssymb}
\usepackage{hyperref}
\usepackage{cite}

\usepackage{cite}

\ifCLASSINFOpdf
\else
\fi
\usepackage{amsmath}
\usepackage{algorithmic}

\usepackage{array}
\usepackage{fixltx2e}
\makeatletter
\let\NAT@parse\undefined
\makeatother
\usepackage{hyperref}  

\begin{document}
\title{Optimal Inter-area Oscillation Damping Control: A Transfer Deep Reinforcement Learning Approach with Switching Control Strategy}
%
\author{\thanks{This work was supported in part by the National Natural Science Foundation of China (Grant No.21773182 (B030103)) and the HPC Platform, Xi'an Jiaotong University.} Siyuan~Liang\thanks{Siyuan Liang (e-mail: siyuan.liang@link.cuhk.edu.hk) is with the Department of Computer Science and Engineering, The Chinese University of Hong Kong, Shatin, 999077, Hong Kong, China.}, Long~Huo, Xin~Chen\thanks{Long Huo (e-mail: eehl921105@stu.xjtu.edu.cn) and Xin Chen (corresponding author, e-mail: xin.chen.nj@stu.xjtu.edu.cn) are with the School of Electrical Engineering, and the Center of Nanomaterials for Renewable Energy, State Key Laboratory of Electrical Insulation and Power Equipment, Xi'an Jiaotong University, Xi'an, 710049, Shaanxi, China.}, and Peiyuan~Sun\thanks{Peiyuan Sun (e-mail: spy2018@stu.xjtu.edu.cn) is with the School of Electrical Engineering, Xi'an Jiaotong University, Xi'an, 710049, Shaanxi, China.}}
\maketitle

\begin{abstract}
Wide-area damping control for inter-area oscillation (IAO) is critical to modern power systems. The recent breakthroughs in deep learning and the broad deployment of phasor measurement units (PMU) promote the development of data-driven IAO damping controllers. In this paper,  the damping control of IAOs is modeled as a Markov Decision Process (MDP) and solved by the proposed Deep Deterministic Policy Gradient (DDPG) based deep reinforcement learning (DRL) approach. The proposed approach optimizes the eigenvalue distribution of the system, which determines the IAO modes in nature. The eigenvalues are evaluated by the data-driven method called dynamic mode decomposition. For a given power system, only a subset of generators selected by participation factors needs to be controlled, alleviating the control and computing burdens. A Switching Control Strategy (SCS) is introduced to improve the transient response of IAOs. Numerical simulations of the IEEE-39 New England power grid model validate the effectiveness and advanced performance of the proposed approach as well as its robustness against communication delays. In addition, we demonstrate the transfer ability of the DRL model trained on the linearized power grid model to provide effective IAO damping control in the non-linear power grid model environment.
\end{abstract}



%

\section{Introduction}
\IEEEPARstart{I}{nter}-area oscillation (IAO) is common in modern interconnected power transmission networks. For example, 15 oscillation events ranging from 0.1 Hz to 0.8 Hz were observed in the China Southern Power Grid between 2008 and 2012 \cite{liu2018effect}. IAO is undesirable and needs to be effectively suppressed for normal power system operations as they limit power transfer capability and can result in system instability. Especially with the rapid growth of stochastic renewable energy resources, IAO damping control becomes more challenging in recent years \cite{eltigani2015challenges}.

Numerous efforts have been put into designing control strategies to dampen IAO. The Power System Stabilizer (PSS) is one of the most commonly used oscillation suppression devices in power systems, which controls the field excitation of a synchronous generator by providing a supplementary feedback signal \cite{devarapalli2022review}. Existing works often use model-based PSS control schemes such as sparsity-promoting linear quadratic control\cite{6740090}, second-order sliding mode-based damping control \cite{liao2016sliding}, delay-dependent state-feedback, dynamic output-feedback method \cite{li2016design} and $etc.$. More model-based IAO damping methods can be found in reviews \cite{zhang2016review, younis2013wide}. For the implementation of model-based IAO damping control, knowledge of the non-linear dynamics or small-signal model is required. However, the exact 
power system model and corresponding parameters may not always be available due to the complex nature of modern power systems, and the time-varying and uncertain operation environment \cite{gupta2021coordinated}.

With the recent breakthroughs in deep learning and broad deployment of phasor measurement units (PMU), Deep Reinforcement Learning (DRL) shows its advanced performances in control and decision-making problems of power systems \cite{chen2022reinforcement}. DRL is a model-free method, only requiring the observable input and output data of the power system. Meanwhile, DRL is robust to the uncertainties of the system, $i.e.$, control signals latency, environment noise, and load variations. A Deep Deterministic Policy Gradient (DDPG) based method and the twin-delayed DDPG method are proposed to overcome various communication delays during damping control \cite{hashmy2020wide, cui2021twin}. To solve the high dimensionality problem of power systems, Mukherjee et al \cite{mukherjee2021scalable} introduce two model reduction approaches for scalable DRL wide-area damping control. For multi-mode oscillation control, a proximal policy optimization (PPO) algorithm for self-tuning of multi-band PSS is proposed \cite{zhang2021novel}. Accounting for $N-1$ and $N-2$ contingencies, a bounded exploratory control-based DDPG (BEC-DDPG) is proposed to coordinate the operation of different PSSs \cite{gupta2021coordinated}. Other related DRL applications include energy storage based oscillation damping via Soft Actor-Critic (SAC) approach \cite{li2021mechanism}, IAO damping using DRL controlled Thyristor Controlled Series Compensator (TCSC) \cite{huang2022damping}, preventing ultra-low-frequency oscillations with asynchronous advantage actor-critic (A3C) algorithm \cite{zhang2020deep}, $etc.$.

For existing DRL-based IAO damping control methods, two challenges remain to be solved. First, how to select the number and location of generators to be controlled are critical for determining the trade-off between control performance and control cost. Selecting all generators in the power system to be controlled under a coordinated manner improves transient response, while high control cost is required \cite{zenelis2018wide}. For a well-studied power system, we can select a specific generator to be controlled by expert experience \cite{9138480} to reduce the control cost. However, expert experience may not always be available. Meanwhile, more generators to be controlled by the DRL-based controller indicated a larger action set needs to be trained during the training process, which increases the computational burden. Second, many DRL-based controllers are designed under the linearized small-signal model of power systems, whose adaptability for practical non-linear power systems remains to be validated.

In this paper, a DRL method and related control strategies are introduced for IAO damping control. The main contributions of this paper are summarized as follows:
\begin{itemize}
	\item A DDPG-based DRL method is proposed to generate wide-area control signals for effectively suppressing IAOs.
	
	\item A subset of generators for the DRL-based control is selected by oscillation mode analysis and participation factors to balance the control performance and control cost.
	
	\item A switching control strategy between the PSSs and DRL controllers is proposed to improve the transient response.
	
	\item The robustness of the proposed DRL method against communication delay is illustrated. In addition, we demonstrate the transfer ability of the DRL model for the non-linear power systems model environment.
\end{itemize}

The remainder of the paper is organized as follows. Section \ref{sec1} introduces the power system model and problem setup of IAO damping control. Section \ref{sec2} presents the DRL method for IAO damping control. Numerical simulations of the IEEE-39 New England power grid model are provided in Section \ref{sec4}. Conclusions are given in Section \ref{sec5}.

\section{Power System Model and Problem Setup}\label{sec1}
In this section, we introduce the power system model for IAO damping control, then we set up the IAO damping control into an optimization problem.
\subsection{Power System Model}\label{sec1-1}
The power system model can be described in a nonlinear, and differential-algebraic form:
\begin{equation}
\boldsymbol{\dot{x}}\left( t \right) =f\left( \boldsymbol{x}\left( t \right) ,\boldsymbol{z}\left( t \right) ,\boldsymbol{u}\left( t \right) ,\boldsymbol{\eta }\left( t \right) \right) 
 \label{eq1}
\end{equation}
\begin{equation}
0=g\left( \boldsymbol{x}\left( t \right) ,\boldsymbol{z}\left( t \right) ,\boldsymbol{u}\left( t \right) ,\boldsymbol{\eta }\left( t \right) \right) 
 \label{eq2}
\end{equation}
where the dynamic and algebraic variables $\boldsymbol{x}(t) \in \mathbb R ^n $ and $\boldsymbol{z}(t) \in \mathbb R ^s$ represent the state, $\boldsymbol{u}(t) \in \mathbb R ^p$ constitutes the control action and $\boldsymbol{\eta}(t) \in \mathbb R ^q$ is white noise. (\ref{eq1}) accounts for the synchronous generators' electromechanical dynamics and the excitation control equipment. (\ref{eq2}) accounts for load flow, generator stator, and power electronic circuit equations. 

Linearizing the power system model in (\ref{eq1}) and (\ref{eq2}) at a stationary operating point and solve the algebraic equations for $\boldsymbol{z}(t)$ will arrive at the linear state-space model:
\begin{equation}
\boldsymbol{\dot{x}}\left( t \right) =\boldsymbol{Ax}\left( t \right)  +\boldsymbol{B}_1\boldsymbol{u}\left( t \right)+\boldsymbol{B}_2\boldsymbol{\eta}\left( t \right)
\label{eq3}
\end{equation}
where $\boldsymbol{A} \in \mathbb R^{n \times n}$, $\boldsymbol{B_1} \in \mathbb R^{n \times p}$, and $\boldsymbol{B_2} \in \mathbb R^{n \times q}$. The system's state variable $\boldsymbol{x}\left( t \right) $ at time $t$ is defined as
\begin{equation}
\begin{cases}
	\boldsymbol{x}\left( t \right) =\left[ \boldsymbol{\theta }\left( t \right) ,\boldsymbol{\omega }\left( t \right) ,\boldsymbol{x}\left( t \right) ^{rem} \right] ^T\\
	\boldsymbol{\theta }\left( t \right) =\left[ \theta _1\left( t \right) ,\theta _2\left( t \right) ,\cdots ,\theta _{n_g}\left( t \right) \right]\\
	\boldsymbol{\omega }\left( t \right) =\left[ \omega _1\left( t \right) ,\omega _2\left( t \right) ,\cdots \omega _{n_g}\left( t \right) \right]\\
	\boldsymbol{x}^{rem}\left( t \right) =\left[ x_{1}^{rem}\left( t \right) ,x_{2}^{rem}\left( t \right) ,\cdots ,x\left( t \right) _{n-n_g}^{rem} \right]\\
\end{cases}
\label{eq4}
\end{equation}
where $n_g$ is the number of generators, $\boldsymbol{\theta }^T,\boldsymbol{\omega }^T\in \mathbb{R}^{n_g}$ are the rotor angles and frequency deviations concerning the normal frequency, and $\boldsymbol{x}_{rem}\in \mathbb{R}^{n-n_g}$ are the remaining state variables.

For the non-linear power system model (\ref{eq1}-\ref{eq2}) or linear power system model (\ref{eq3}), the IAO damping controller can be described as \cite{6740090}: 
\begin{equation}
\boldsymbol{u}\left( t \right) =\boldsymbol{u}_{loc}\left( t \right) +\boldsymbol{u}_{wac}\left( t \right)  
\label{eq5}
\end{equation}

In (\ref{eq5}), the first term is the local control signal $\boldsymbol{u}_{loc}(t)$ from the PSS on each generator, which is designed based on locally measured states. The second term is the wide-area control signal $\boldsymbol{u}_{wac}(t)$ from the DRL controller, which is an additional control signal designed based on integrating information from all generators. In practice, signals from the controller (\ref{eq5}) will adjust the voltages applied at the field windings of the controlled generators. For generator $i$, the local control signal of the PSS in the Laplace domain denotes \cite{2009.0669}:
\begin{equation}
u_{loc,i}(s)=k_i \cdot \frac{T_{w,i}s}{1+T_{w,i}s} \cdot \frac{1+T_{n1,i}s}{1+T_{d1,i}s} \cdot \frac{1+T_{n2,i}s}{1+T_{d2,i}s} \cdot \dot \theta_i(s)
\label{eq6}
\end{equation}
where $\theta_i(s)$ is the rotor angle of generator $i$ and $k_i$, $T_{w,i}$, $T_{n1,i}$, $T_{d1,i}$, $T_{n2,i}$, and $T_{d2,i}$ are controller gains of the PSS.

For wide-area control $\boldsymbol{u}_{wac}(t)$, a feedback control mechanism is introduced:
\begin{equation}
\boldsymbol{u}_{wac}\left( t \right) =-\boldsymbol{Kx}\left( t \right) 
\label{eq7}
\end{equation} 
where $\boldsymbol{K}\in \mathbb{R}^{p\times n}$ is the feedback gain matrix. 
\subsection{Problem Setup}\label{sec1-3}
The physical nature of power system oscillations is determined by the spectrum of close-loop system matrix $\boldsymbol{A}-\boldsymbol{B}_1\boldsymbol{K}$, $i.e.$, the eigenvalues of $\boldsymbol{A}-\boldsymbol{B}_1\boldsymbol{K}$ reveal the natural frequency and damping ratio of each oscillation mode:
\begin{equation}
f_i=\sqrt{\mathrm{Re}\left( \lambda _i \right) ^2+\mathrm{Im}\left( \lambda _i \right) ^2}
\label{naturalf}
\end{equation}
\begin{equation}
\zeta _i=-\frac{\mathrm{Re}\left( \lambda _i \right)}{f_i}
\label{dampingr}
\end{equation}	
where $\lambda _i$ is the $i$th pair eigenvalues of matrix $\boldsymbol{A}-\boldsymbol{B}_1\boldsymbol{K}$, $\mathrm{Re}\left( \cdot \right)$ and $\mathrm{Im}\left( \cdot \right)$ denote the real and imaginary parts of $\lambda _i$, respectively. (\ref{naturalf}) is the natural frequency of $i$th oscillation mode, describing how fast the motion oscillates. (\ref{dampingr}) is the damping ratio of $i$th oscillation mode, describing how much amplitude decays per oscillation. Accordingly, we set the problem of IAO damping control into a framework for the optimization of eigenvalues distributions:

\begin{subequations}
\begin{align}
\begin{aligned}	
\mathrm{minimize}\,\,J\left( \boldsymbol{K}(t) \right) &= \alpha\sum_i{\left[ \mathrm{Re}\left( \lambda _i \right) ^2-\mathrm{Re}\left( \hat{\lambda}_i \right) ^2 \right]}\\
&+\beta\sum_i{\mathrm{Im}\left( \lambda _i \right) ^2}
\label{eq8a}
\end{aligned}
\end{align}
\,\,\,\,\,\,\,    \text{subject to}
\begin{align}
\text{dynamics:}\,\,\,\,\, &\boldsymbol{\dot{x}}(t)=\boldsymbol{Ax}(t)+\boldsymbol{B}_1\boldsymbol{u}(t)+\boldsymbol{B}_2\boldsymbol{\eta}\left( t \right)
\label{eq8b}
\\
\text{stability:}\,\,\,\,\,\,\,\, &(\boldsymbol{A}-\boldsymbol{B}_1\boldsymbol{K}(t)) \, \text{Hurwitz}
\label{eq8e}
\end{align}

\label{eq8}
\end{subequations}	
where $\hat{\lambda}_i$ is the $i$th pair eigenvalues of the open-loop matrix $\boldsymbol{A}$, $\alpha$ and $\beta$ are weight constants. As shown in the objective (\ref{eq8a}), the goal is to minimize the real part differences between eigenvalues of close-loop matrix $\boldsymbol{A}-\boldsymbol{B}_1\boldsymbol{K}$ and eigenvalues of open-loop matrix $\boldsymbol{A}$, as well as the imaginary part of eigenvalues of close-loop matrix $\boldsymbol{A}-\boldsymbol{B}_1\boldsymbol{K}$. Since from (\ref{naturalf}) and (\ref{dampingr}) we can find that when the real part of eigenvalues does not change significantly and the imaginary part of eigenvalues are nearly zeros, the natural frequencies will decrease and the damping ratios will increase. Smaller natural frequencies and larger damping ratios make the oscillations of the close-loop system become slower and amplitude decay per oscillation becomes larger. In this paper, the control gain matrix $\boldsymbol{K}(t)$ serves as the variable to be optimized with the constraints of power system dynamics (\ref{eq8b}), stability criterion (\ref{eq8e}) and implicit constrain of linear feedback control (\ref{eq5}) and (\ref{eq6}). Note that $\boldsymbol{K}(t)$ is time-varying during the damping control.

The optimization problem in (\ref{eq8}) can be further formatted as a Markov Decision Process (MDP), which converts the damping control of IAOs into sequence decision making. The MDP can be represented as a five-tuple$(S, A, P, R, \gamma)$, among which $S$ is the state of the system, $A$ is the action set, $P$ is the state transition probability, $R$ represents the reward for each transition, and $\gamma$ is a discount factor \cite{sutton2018reinforcement}. At time step $t$, the MDP yields:

\textbf{System State:}
The system state corresponds to the power system state variable $\boldsymbol{s}_t=\boldsymbol{x}\left( t \right)$.

\textbf{Action:}
The immediate action is denoted as $\boldsymbol{a}_t = \boldsymbol{K}(t)$.

\textbf{Reward:} The reward takes the negative value of the objective (\ref{eq8a}), $r_t=-J\left( \boldsymbol{K}\left( t \right) \right)$. The goal of the MDP is to find the optimal action $\boldsymbol{a}_t$ at each step, based on which the largest expect reward can be obtained accordingly.

\textbf{State Transition:}
The state transition at time $t$ is a tuple $(\boldsymbol{s}_t,\boldsymbol{a}_t,\boldsymbol{s}_{t+1},r_t)$. The next system state $\boldsymbol{s}_{t+1}$ is determined by the action $\boldsymbol{a}_t$, system state $\boldsymbol{s}_t$, and the state transition probability $\boldsymbol{s}_{t+1}=P(\boldsymbol{s}_t,\boldsymbol{a}_t)$. 

\section{DRL Method}\label{sec2}
To solve the MDP introduced in the former section, a model-free DRL method based on the Deep Deterministic Policy Gradient (DDPG) model is used. In the following, we introduce the architecture, performance improvement techniques, and training of DDPG.
\begin{figure*}[h]
  \centering
  \includegraphics[width=6.6in]{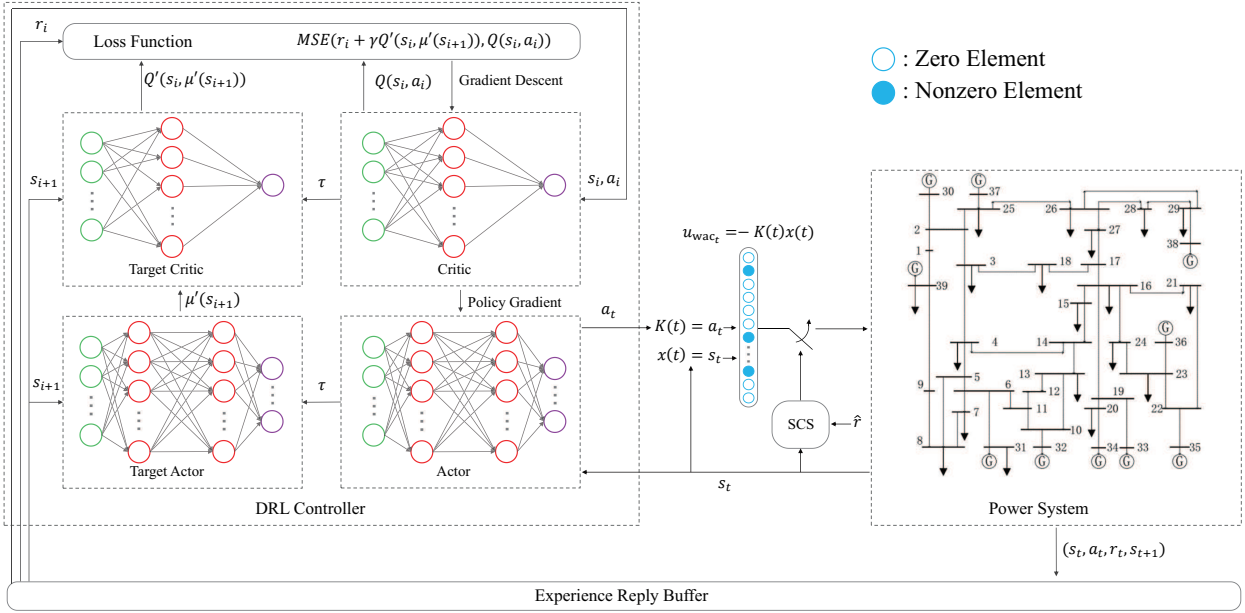}
  \caption{The architecture of DDPG for IAO damping control. The circles in green, red, and blue represent neurons in the input layers, hidden layers, and output layers, respectively. }
  \label{DDPG}
  \end{figure*}

\subsection{Architecture}
Fig. \ref{DDPG} illustrates the architecture of the DDPG. The DDPG includes two main components interacting with each other: the DRL controller and environment. The DRL controller is composed of the actor network, critic network, and target network. The environment corresponds to the power system. Details of each component are given as follows.

\textbf{Actor network:}
The actor-network is a multi-layer fully-connected neural network mapping the system state $\boldsymbol{s}_t$ to the action $\boldsymbol{a}_t$ by a non-linear actor function $\mu$: 
\begin{equation}
	\boldsymbol{a}_t=\mu(\boldsymbol{s}_t)
	\label{eq10}
\end{equation}

Specifically, the actor-network includes three layers: the input layer, the hidden layer, and the output layer. The input of the actor-network is system state $s_t$. The hidden layer $\boldsymbol{h_{a}}$ can be described as:
\begin{equation}
\boldsymbol{h_{a}}=relu(\boldsymbol{W_{ha}}\cdot \boldsymbol{s}_t^T+\boldsymbol{b_{ha}})
\label{eq12}
\end{equation}
where $\boldsymbol{W_{ha}}$, $\boldsymbol{b_{ha}}$, and $relu$ function are weights, biases, and the activation function of the hidden layer, respectively. The $relu$ function reads:
\begin{equation}
relu(x)
\begin{cases}
\ 0 &\ \ \ \ if\ \ x < 0\\
\ x &\ \ \ \ if\ \ x\geqslant0
\end{cases}
\label{eq13}
\end{equation}

The output of the actor-network is action $\boldsymbol{a}_t$:
\begin{equation}
\boldsymbol{a}_t=tanh(\boldsymbol{W_{oa}}\cdot \boldsymbol{h_a}+\boldsymbol{b_{oa}})
\label{eq14}
\end{equation}
where $\boldsymbol{W_{oa}}$, $\boldsymbol{b_{oa}}$, and hyperbolic tangent function $tanh$ are weights, biases, and the activation function of the output layer.

\textbf{Critic network:}
The critic network is a multi-layer fully-connected neural network aiming to estimate the expected reward of the current action and state:
\begin{equation}
	Q^\mu(\boldsymbol{s}_t,\boldsymbol{a}_t)=E_\mu\left[\sum_{k=0}^\infty\gamma^k\cdot r_{t+k} \Big|\boldsymbol{s}_t,\boldsymbol{a}_t\right]
	\label{eq11}
\end{equation}
where $Q^\mu$ is the critic function, $0<\gamma<1$ is the discount factor accounting for the future and immediate rewards, and $E_{\mu}\left[ \cdot \right]$ represents the expected value.

The critic network has three layers: the input layer, the hidden layer, and the output layer. 
The input of the critic network includes the system state $\boldsymbol{s}_t$ and the action $\boldsymbol{a}_t$. The hidden layer $\boldsymbol{h_c}$ can be described as:
\begin{equation}
\boldsymbol{h_c}=relu(\boldsymbol{W_{hc}}\cdot (\boldsymbol{s}_t,\boldsymbol{a}_t)^T+\boldsymbol{b_{hc}})
\end{equation}
where $\boldsymbol{W_{hc}}$, $\boldsymbol{b_{hc}}$, and the $relu$ function are weights, biases, and the activation function of the hidden layer.

The output of the critic network is the expected reward $Q_t$ according to the value from the output layer:
\begin{equation}
Q_t=\boldsymbol{W_{oc}}\cdot \boldsymbol{h_c}+\boldsymbol{b_{oc}}
\end{equation}
where $\boldsymbol{W_{oc}}$ and $\boldsymbol{b_{oc}}$ are weights and biases of the output layer. Note that no activation function is used in the output layer of the critic network.

\textbf{Target network:}
Since we also use the critic network to estimate the expected reward, the updating process is prone to divergence, which is undesired. Inspired by \cite{mnih2015human}, one possible solution is creating the target actor and critic networks which are expressed as $\mu'$ and $Q'$ respectively. The parameters of actor and critic networks are donated as $\Theta$, and the parameters of target actor network and target critic networks are donated as $\Theta'$. The two target networks are updated by tracking the learned networks $\Theta'\leftarrow \tau \Theta+(1-\tau) \Theta'$ while $\tau \ll 1$. Thus, the target networks are updated more slowly, which provides steady support for actor and critic networks and then improves the stability of updating significantly \cite{8882242}.

\textbf{Environment:} The power system acts as the environment for DDPG architecture. Phasor Measurement Units (PMUs) are assumed deployed at each generator bus, making the data of rotor angle and frequency of each generator available. The power system provides the data of rotor angles and frequencies as the system states $\boldsymbol{s}_t$ for the DRL controller. In turn, the DRL controller sends the action values $\boldsymbol{a}_t$ as wide-area control signals to the power system.

\begin{figure}[h]
  \centering
  \subfigure[Generator frequency components]{
    \includegraphics[width=0.22\textwidth]{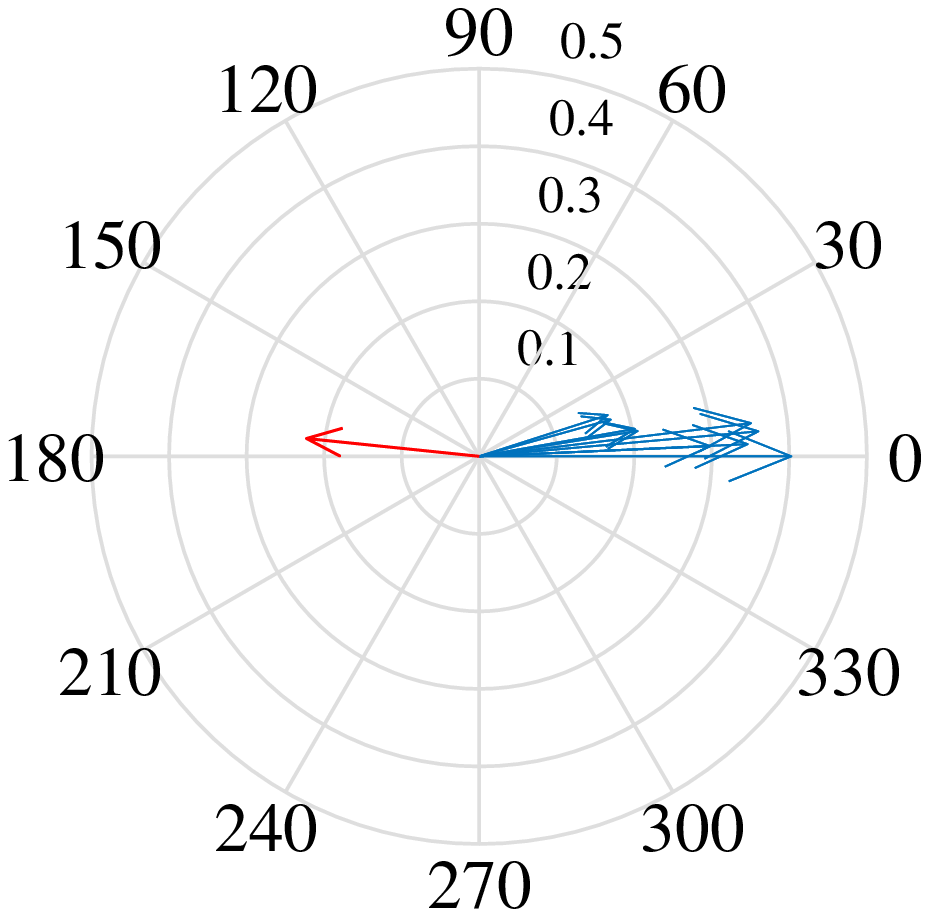}\label{fig_c5a}
  }
  \subfigure[Participation factors]{
    \includegraphics[width=0.22\textwidth]{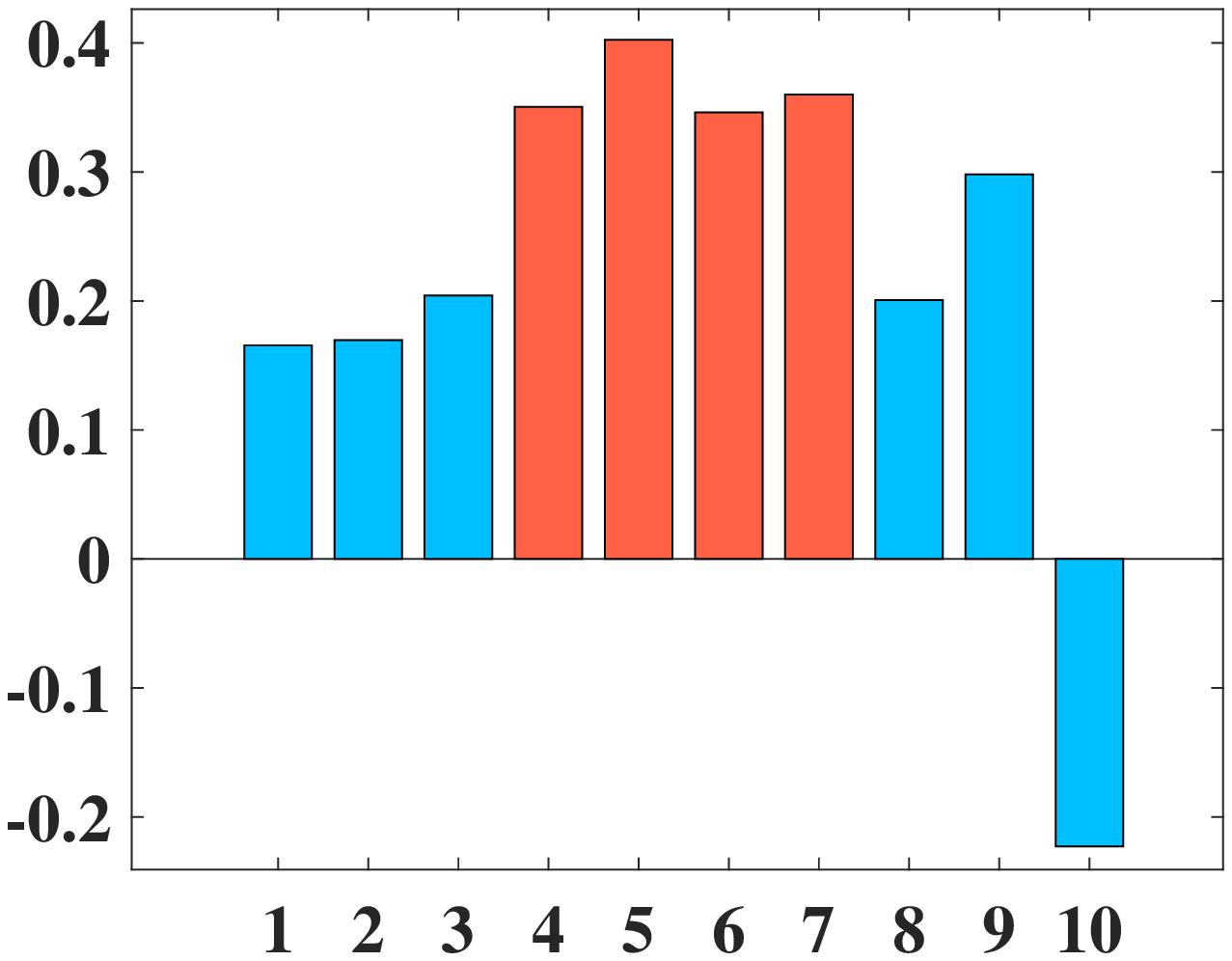}\label{fig_c5b}
  }
  \caption{Example of the IAO mode and corresponding participation factors.}
  \label{fig_c5}
  \end{figure}

\subsection{Performance Improvement Techniques}\label{sec2-2}

\textbf{Reward:} To ensure the stability constraint in (\ref{eq8e}), we check the maximum eigenvalue $\lambda_{\max}(\boldsymbol{A}-\boldsymbol{B}_1\boldsymbol{K}(t))$ of the close-loop system matrix for the linear state space model (\ref{eq3}) with the IAO damping controller (\ref{eq4} to \ref{eq5}), $i.e.$, the system is unstable when $\lambda_{\max}(\boldsymbol{A}-\boldsymbol{B}_1\boldsymbol{K}(t)) > 0$; otherwise, the system is stable. During the training of DRL, we penalize the instability by a large negative reward $r_t = -300$.

It should be noted that, for the DRL controller, the power system is a black box model, $i.e.$, the eigenvalues shown in objective \ref{eq8a} and the maximum eigenvalue used for check the stability constraint \ref{eq8e}, can not be directly calculated by the DRL controller, while the system states, $i.e.$, the generator frequency deviations and rotor angles, can be observed by the DRL controller. Hence, a data-driven algorithm called dynamic
mode decomposition (DMD) \cite{schmid2010dynamic} is applied for eigenvalues estimation based on the observable system states. The DMD algorithm can be described as follows \cite{tu2013dynamic}:
\begin{itemize}
	\item \textbf{Step 1:} Build the information matrices from the PMU measurements:
	\begin{subequations}
		\begin{align}
			&\boldsymbol{X}\triangleq \left[ \boldsymbol{x}(0),\boldsymbol{x}(1)\cdots ,\boldsymbol{x}(T-1) \right] 
			\\
			&\boldsymbol{Y}\triangleq \left[ \boldsymbol{x}(1),\boldsymbol{x}(2)\cdots ,\boldsymbol{x}(T) \right] 
		\end{align}		
	\end{subequations}	
	where vectors $\boldsymbol{x}(t)$, $t=0,1,\cdots T$ denote the observable input and output system states defined in (\ref{eq4}).
	\item \textbf{Step 2:} Compute the singular value decomposition of matrix $\boldsymbol{X}$ as
	\begin{equation}
		\boldsymbol{X}=\boldsymbol{U\varSigma V}^*
	\end{equation}
	where $\boldsymbol{U}$, $\boldsymbol{V}$, and  $\boldsymbol{\varSigma}$ is the matrix of left singular vectors, right singular vectors, and singular values, respectively, and $\left( \cdot \right) ^*$ represents the transposition operation.
	\item \textbf{Step 3:} Define the matrix
	\begin{equation}
		\boldsymbol{F}\triangleq \boldsymbol{U}^*\boldsymbol{YV\varSigma }^{-1}
	\end{equation}
	\item \textbf{Step 4:} Compute the approximate eigenvalues and eigenvectors by
	\begin{subequations}
		\begin{align}
			&\boldsymbol{F\phi }_i=\tilde{\lambda}_i\boldsymbol{\phi }_i
			\\
			&\boldsymbol{\tilde{\varphi}}_i=\boldsymbol{U\phi }_i
		\end{align}		
	\end{subequations}
	where $\tilde{\lambda}_i$ and $\boldsymbol{\tilde{\varphi}}_i$ represent the estimated $i$th eigenvalue and the corresponding eigenvector.
\end{itemize}

\begin{figure}[h]
	\centering
	\subfigure[Real part AEs]{
		\includegraphics[width=0.22\textwidth]{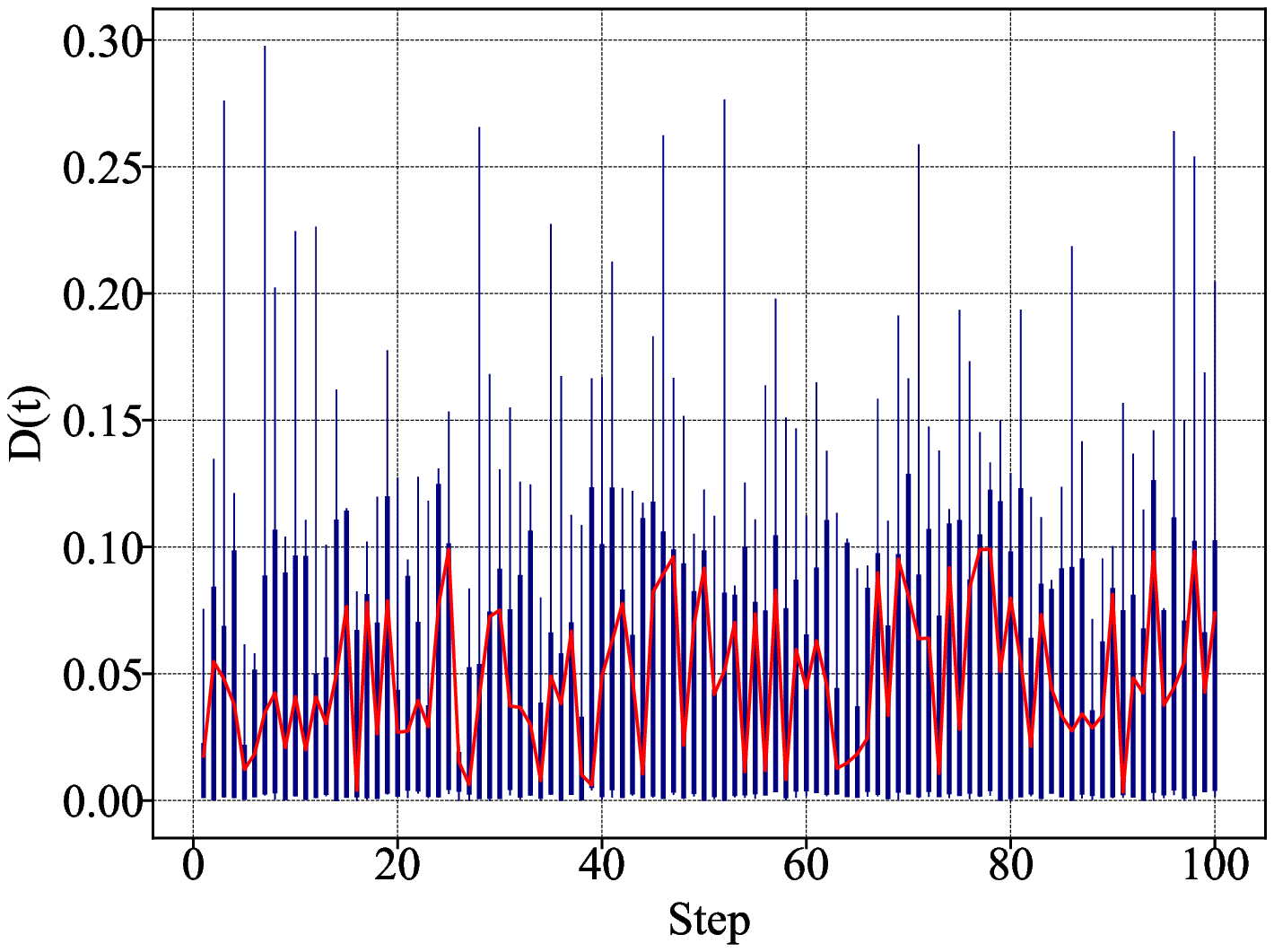}\label{fig_real}
	}
	\subfigure[Imaginary part AEs]{
		\includegraphics[width=0.22\textwidth]{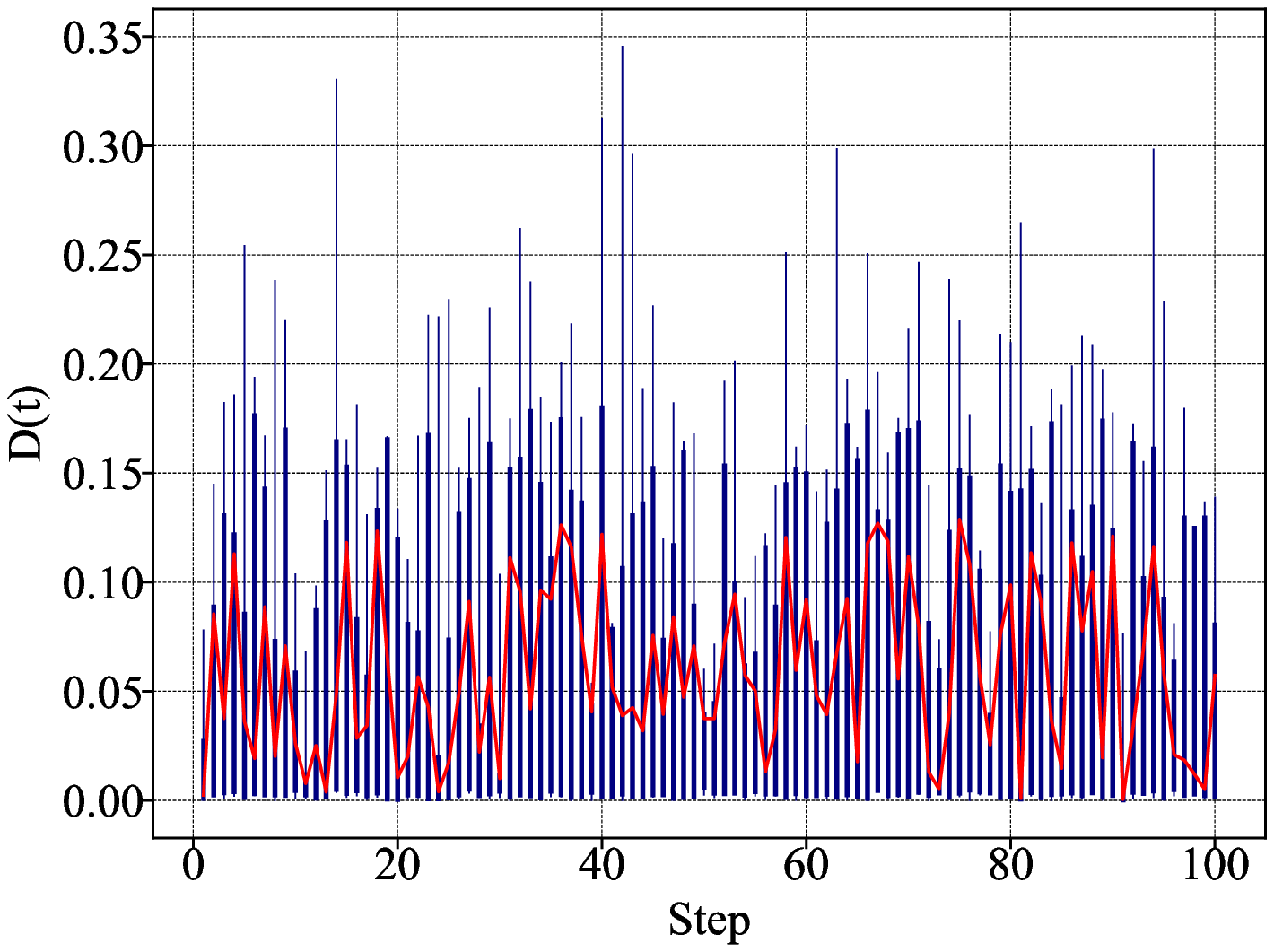}\label{fig_imag}
	}
	\caption{Absolute Errors (AEs) between exact eigenvalues and estimated eigenvalues. (a) shows AEs of the real part and (b) shows AEs of the imaginary part. The lines in blue and red denote the AEs every time step and the average AEs, respectively.}
	\label{fig_estimated}
\end{figure}

As shown in Fig. \ref{fig_estimated}, the small Absolute Errors (AEs) between exact eigenvalues and estimated eigenvalues of the IEEE-39 New England power grid model. The AEs is calculated during an oscillation damping control simulation. The average AEs for the real part and the imaginary part of eigenvalues are below 0.10 and 0.13, respectively.

\textbf{Training Reward:}
During the training of DRL, usually experience transitions were uniformly
sampled from the experience replay buffer. However, this approach simply replays transitions at the same frequency that they were originally experienced regardless of their significance, which leads to inefficient training. Hence, the Prioritized Experience Replay (PER) \cite{schaul2015prioritized} is used in this paper. Since it is the unideal experience that can bring more significant adjustment to the agent during training, the PER defines the unideal experience as the prioritized target. The prioritized target is automatically obtained every $M$ episodes from the proposed test, so that the agent could try repeatedly on the unideal circumstance with its updating DNNs until the transition become ideal, or in other words, the agent masters the task that it once failed.


\section{Control Strategies}\label{sec3}
\subsection{Generator Selection}\label{sec3-1}
To save the control cost, not all generators, but a subset of generators are selected to be controlled by the DRL controller. Participation factors are used to select the subset of generators. Participation factors are derived from eigenvalues and eigenvectors of the open-loop matrix $\boldsymbol{A}$, illustrating the impact of each generator on a given oscillation mode. 

As an example, Fig. \ref{fig_c5} shows an oscillation mode and the corresponding participation factors of the IEEE-39 New England power grid model with ten generators. The IEEE-39 New England power grid model is displayed in Fig. \ref{DDPG}. We denote the generators at bus 30 to 39 as generator 1 to generator 10. In Fig. \ref{fig_c5a}, the oscillation mode is an IAO mode, where generator 10 is swinging against other generators. Fig. \ref{fig_c5b} reflects that generators 4 to generator 7 have higher impact on the IAO mode due to their larger participation factors, and the participation factors of the rest generators are relatively smaller. Hence, for the given IAO mode shown in Fig. \ref{fig_c5a}, generators 4 to generator 7 are selected to be controlled by the DRL controller.

\subsection{Switching Control Strategy}\label{sec3-2}

Recalling (\ref{eq5}), the IAO damping controller has two types of control signals. One is the local control signal from the PSS on each generator, damping the oscillation with information from the individual generator. The other is the wide-area control signal from the DRL controller, damping the oscillation with integrated information from all generators, $i.e.$, the DRL controller generates wide-area control signals based on the rotor angles and frequency deviations of all generators in the power system. In this section, a Switching Control Strategy (SCS) is proposed to make the DRL controller participate in the IAO damping control Intermittently. As shown in Fig. \ref{DDPG}, the SCS is used to decide whether the DRL controller participates in the IAO damping control according to the reward $r_t$. The rule of SCS is shown below:

\begin{equation}
	\left\{
	\begin{array}{lc}
		\boldsymbol{u}(t)= \boldsymbol{u}_{loc}(t)+\boldsymbol{u}_{wac}(t),& P(t)> \hat{r} \\
		\boldsymbol{u}(t)=\boldsymbol{u}_{loc}(t),& P(t)\le \hat{r}\\
	\end{array}
	\right.
	\label{scsrule}
\end{equation}

\begin{equation}
\begin{aligned}
P\left( t \right) &=\kappa _1\sum_{i=1}^{n_g}{\sum_{j=1}^{n_g}{\left[ \omega _i\left( t \right) -\omega _j\left( t \right) \right] ^2}}+\kappa _2\sum_{i=1}^{n_g-1}{\omega _{i}^{2}\left( t \right)}
\\
&+\kappa _3\sum_{i=1}^{n_g-1}{\left[ \theta _i\left( t \right) -\theta _{ref}\left( t \right) \right] ^2}
\end{aligned}
\label{energy}
\end{equation}
where $\hat{r} > 0$ is a given threshold for switching the DRL controller on or off, $\kappa _1, \kappa _2$and $\kappa _3$ are weights, $\theta _{ref}$ is the phase of the reference generator. $P(t)$ is a system energy-like framework, by using which we quantify the control performances.  In (\ref{energy}), the first term quantifies the amplitude of IAOs between each pair of generators, the second and third terms quantify the kinetic and potential energy of generators, respectively. The smaller $P(t)$ is, the better the the performance of transient responses. Specifically, at the beginning of IAO, $P(t)> \hat{r}$ will be satisfied concerning relatively large frequency violation and angular difference. The wide-area controller is switched on for effective suppression of the oscillation, $i.e.$, the control signals from the trained DRL model provide auxiliary effort on fast oscillation damping. However, the wide-area controller requires additional cost for remote communication, since the DRL needs to obtain the global system states. Considering communication cost, when the reward $P(t) \le \hat{r}$, which means the oscillation has been considerably suppressed, the wide-area controller is switched off and no remote communication cost is needed.

\section{Numerical Simulations}\label{sec4}
In this section, the IEEE-39 New England power grid model is used to validate the proposed method. The diagram of the IEEE-39 New England power grid model is illustrated on the right side of Fig. \ref{DDPG}. The power grid model includes 39 buses and 10 generators. The corresponding electrical parameters are obtained from the power system toolbox \cite{207380}. We denote the generators at bus 30 to 39 as generator 1 to generator 10.

The DRL model introduced in Section \ref{sec2} is trained on the Intel(R) Xeon(R) CPU E5-2620 v3 CPU platform with the TensorFlow software environment. During the training, the minibatch size is 32, and the capacity of the replay buffer is 1000. The max episodes and max steps in one episode are set as 5000 and 500, respectively. The discount factor $\gamma$ is 0.95. The learning rate for both actor and critic networks is 0.0001. The white noise $\boldsymbol{\eta}(t)$ is sampled from Gaussian distribution with zero mean and standard deviation 0.01.

\begin{figure}[htb]
	\centering
	\includegraphics[width=3in]{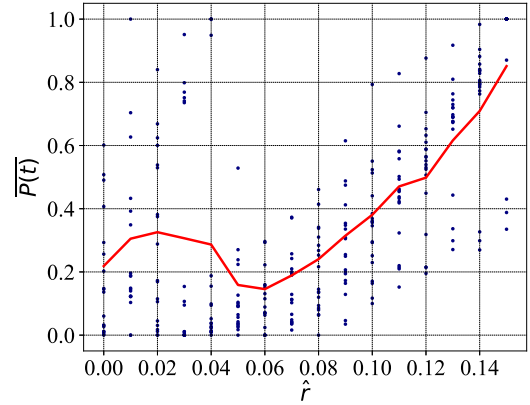}
	\caption{The normalized cumulative system energy-like framework $\bar{P}(t)$ with different threshold $\hat{r}$. Each blue point represents the normalized $\bar{P}(t)$ for one control. The red curve is the average normalized $\bar{P}(t)$ of 20 different controls using the same threshold $\hat{r}$.}
	\label{SCS}
\end{figure}

For SCS in Section \ref{sec3-2}, we made use of the cumulative system energy-like framework to determine suitable the threshold $\hat{r}$. The cumulative system energy-like framework $\bar{P}(t)$, $a.k.a$, the return, is is the sum of $P(t)$ values at each time point in a control process and the goal of control is to minimize its value. Different threshold $\hat{r}$ gives different switching points between the wide-area DRL controller and the local PSS controller, and in turn, leads to different $\bar{P}(t)$. Fig. \ref{SCS} shows the normalized $\bar{P}(t)$ changes with the different threshold $\hat{r}$. The normalized $\bar{P}(t)$ is obtained by the max-min normalization to characterize the $\bar{P}(t)$ concerning different thresholds on an equal footing. Each blue point represents a normalized $\bar{P}(t)$ of one control process with a given $\hat{r}$. At each threshold, we do controls with 20 different initial states, and the average normalized $\bar{P}(t)$ of the 20 times control is shown by the red curve. The average normalized $\bar{P}(t)$ has a minimum at $\hat{r}\approx 0.06$ and then increases monotonically. Since the control of the proposed model aims to minimize the $\bar{P}(t)$, $\hat{r}=0.06$ is selected as the reward threshold for the SCS rule in (\ref{scsrule}).

\subsection{Performance Evaluation}
As shown in Section \ref{sec3-1}, the performance of the trained DRL model is tested under the IAO mode that generator 1 to generator 9 oscillates against generator 10. According to Section \ref{sec3-1}, the DRL controllers are deployed on a subset of generators including generator 4 to generator 7. The initial conditions are aligned with the eigenvector of the corresponding IAO mode. Generator 10 is set as a reference generator without the damping controller. We compared the proposed DRL method with the conventional local PSSs method. For the local PSS in (\ref{eq6}), the control gains are chosen according to the tuning strategy \cite{2009.0669}.

\begin{figure}[h]
	\centering
	\includegraphics[width=3.8in]{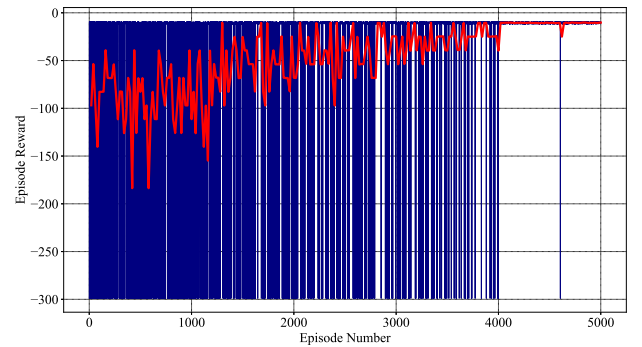}
	\caption{The learning curve of the DRL model during training. The line in blue shows the reward of each episode. The line in red shows the average reward for every 20 episodes.}
	\label{traincurve}
\end{figure}

Fig. \ref{traincurve} shows the learning curve of the DRL model during training. The line in blue shows the average reward for each episode. The line in red shows the average reward for every 20 episodes. The rewards in blue approaching -300 indicate the penalty for unstable states during training. After 4000 episodes, nearly no unstable states exist. The average reward in red starts at a low value and gradually increases. Finally, the average reward keeps a high value close to zero after 4000 episodes, validating the effectiveness of the training process.

\begin{figure}[h]
	\centering
	\includegraphics[width=3.3in]{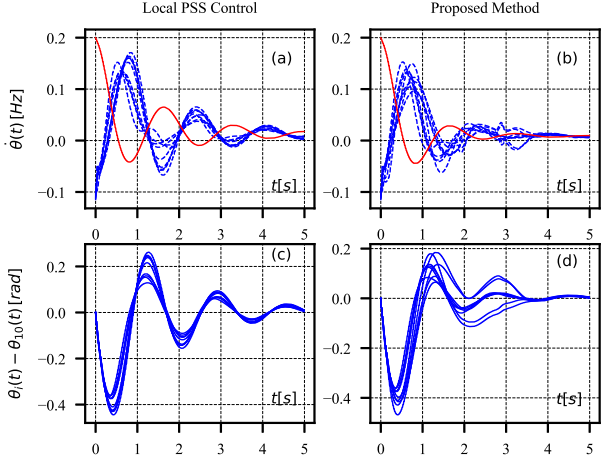}
	\caption{Time-domain simulation of the linearized state-space model (\ref{eq3}) of the IEEE-39 New England power grid model. The subfigures show the generator frequency deviations and the angle differences resulting from the use of the local PSS control and the proposed method.}
	\label{linear}
\end{figure}

Fig. \ref{linear} illustrates the time-domain control results of the linearized state-space model (\ref{eq3}) of the IEEE-39 New England power grid model. The first and second row shows the generator frequency deviations and angle differences, respectively. The angle differences reflect the inter-area power transfer, $i.e.$, larger angle differences correspond to larger inter-area power transfer. The first and second column is control results under local PSS control and the proposed DRL method, respectively. For generator frequency deviations, The lines in blue are frequency deviations of generator 1 to generator 9, and the lines in red are frequency deviations of generator 10. The IAO is shown by generator 10 oscillates against other generators. It can be found that compared with local PSS control, the frequency differences between generator 10 and the other generators are smaller, and the frequency deviation of all ten generators decays faster under the control of the proposed method. Although the angle differences of the proposed method are slightly larger around 2s to 4s, they are more effectively suppressed after 4s of the proposed method.

\begin{figure}[b]
	\centering
	\includegraphics[width=3.3in]{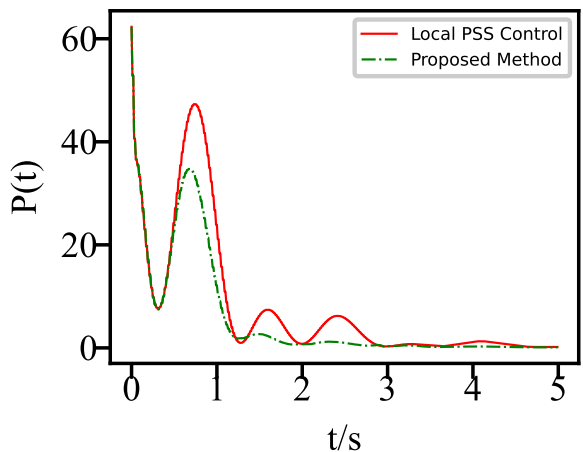}
	\caption{The quantified control performance $P(t)$ of different control methods for the linear power system environment (\ref{eq3}).}
	\label{per}
\end{figure}

Fig. \ref{per} displays the $P(t)$ during the IAO damping control. The solid line in red and the dashed line in green are the $P(t)$ with local PSS control and the proposed DRL method, respectively. At $t=0$ of initial perturbation of IAO, $P(t)$ is the largest. Then the $P(t)$ gradually decays due to the frequency deviations and angle differences are suppressed by the IAO damping control. However, for local PSS control and the proposed DRL method, the decay speed and amplitude per oscillation are different. It can be found that the $P(t)$ under the proposed method decays faster with a larger decay amplitude per oscillation. Hence, from the quantified performance given by the $P(t)$, the proposed method shows advanced control performance.

\subsection{Transfer ability for the nonlinear power system}
\begin{figure}[t]
	\centering
	\includegraphics[width=3.3in]{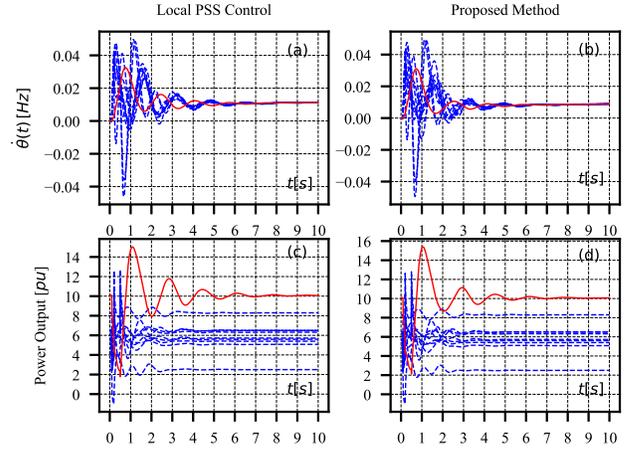}
	\caption{Time-domain simulation of the full nonlinear differential-algebraic power system environment (\ref{eq1} to \ref{eq2}) of the IEEE-39 New England power grid model. The subfigures show the generator frequency deviations and active power outputs using the local PSS control and our proposed method both without delay and with an 800 ms delay. Initially, the system is at a steady state, and a three-phase fault is triggered at line \{3,4\} at 0.1 s, which is cleared at 0.2 s and the remote end is cleared at 0.5s.}
	\label{nonlinear}
\end{figure}

\begin{figure}[b]
	\centering
	\includegraphics[width=3.3in]{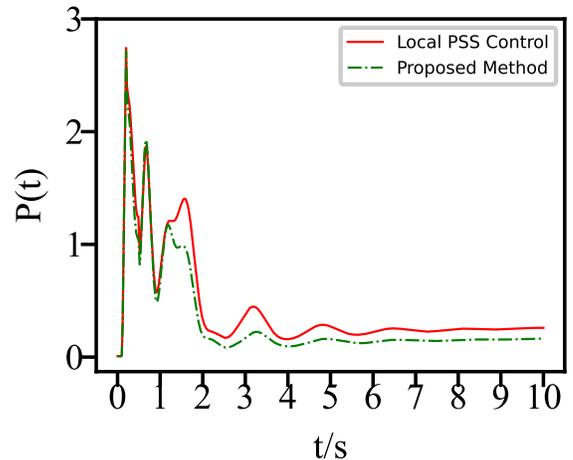}
	\caption{The quantified control performance $P(t)$ of different control methods for the full nonlinear differential-algebraic power system environment (\ref{eq1}-\ref{eq2}).}
	\label{quant}
\end{figure}

In this section, we demonstrate the transfer ability of the proposed DRL method for the more practical nonlinear power system environment. We directly apply the proposed DRL model trained in the linear power system environment (\ref{eq3}) to control the IAO in the full nonlinear differential-algebraic power system environment (\ref{eq1}-\ref{eq2}). It is meaningful to validate the transfer ability of the DRL model applied in the non-linear power system environment. Since in practical power systems, the dynamics are always non-linear. Fig. \ref{nonlinear} shows the control results for the corresponding nonlinear power system. The first and second column is the control result under local PSS control and our proposed method, respectively. The first and second row is the frequency deviations and power outputs, respectively. For frequency deviations, the curves in red correspond to generator 10 and the curves in blue correspond to other generators. A three-phase fault is triggered at line \{3,4\} at 0.1s, which is cleared at 0.2s and the remote end is cleared at 0.5s. It can be found from Fig. \ref{nonlinear} when the DRL model transfers to the non-linear power system environment, it is also effective for damping the IAO. Specifically, compared with the local PSS control, the proposed DRL method suppress the IAO faster and allows the power system to reach a new steady state earlier.

\subsection{Performance Under time Delays}
\begin{figure}[H]
	\centering
	\includegraphics[width=3.3in]{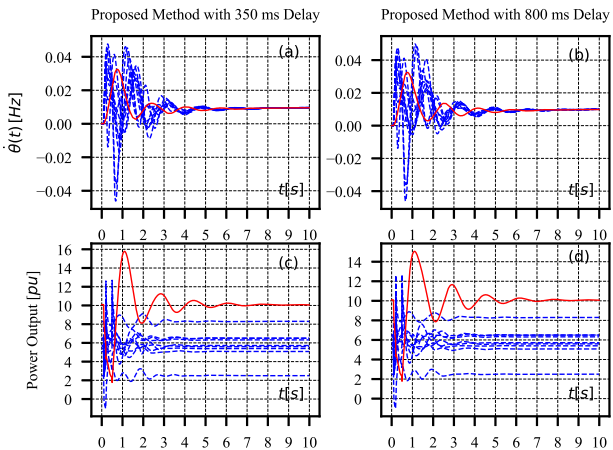}
	\caption{The control performance of different control methods under communication delay. We set the time delay as 350 and 800 ms respectively.}
	\label{delay}
\end{figure}
In this section, we further investigate the robustness of the proposed method against communication delays. In practical, wide-area control is inevitable to face the challenge of time delays. The time delays may arise from different resources such as the long-distance communication channels, latencies and multiple data rates in the SCADA (supervisory control and data acquisition) network, asynchronous measurements, and local processing and operating times \cite{6740090}. Typically, time delays of communication in power systems range from 100ms (fiber-optic cables) to 700ms (satellite link) \cite{9138480}. We set a moderate time delay as 350ms and a  severe time delay as 800ms to test the robustness of the proposed method. Fig. \ref{delay} shows the control performance of the proposed method. Comparing Fig. \ref{nonlinear} (b) and (d) with Fig. \ref{delay}, we observe that the close-loop performance of the proposed method is not affected significantly by 350 and 800ms delays, which ensures the robustness of the proposed method against time delays.

\section{Conclusions}\label{sec5}
A transfer deep reinforcement learning approach with switch control strategy is developed for the optimal inter-area oscillation damping control. Compared with the local PSS control, the faster decay speed of inter-area oscillations and the larger decay amplitude per oscillation can be realized by the control of proposed DRL method. The switch control strategy between the local PSS and DRL controller helps to improve the performance as well as reduce the communication cost. We illustrate that not all generators, but a subset of generators selected by the participation factors needs to be controlled, mitigating the computing burden of the DRL controller.
Considering the practical application, we show that the proposed DRL method is robustness against the communication delays. The performance is slightly drop even for the serve 800ms time delays. In addition, we demonstrate the transfer ability of the proposed method for full nonlinear differential-algebraic power system environment. The DRL model trained in the linear power system environment can be directly transferred to suppress the inter-area oscillation in nonlinear power system environment.


%



\section*{ACKNOWLEDGMENT}

The authors would like to thank Prof. Florian Dorfler and Dr. Xiaofan Wu for helpful discussions regarding the Power Systems Toolbox.

\ifCLASSOPTIONcaptionsoff
  \newpage
\fi
\bibliographystyle{IEEEtran}
\bibliography{eecs}\ 

\end{document}